**A Quantitative Method for Estimating the Human Development Stages by Based on the Health State Function Theory and the Resulting Deterioration Process**

**Christos H Skiadas[1] and Charilaos Skiadas[2]**

[1]Technical University of Crete, Chania, Crete, Greece
 E-mail: skiadas@cmsim.net
[2]Hanover College, Indiana, USA
 E-mail: skiadas@hanover.edu

**Abstract:** The Health State Function theory is applied to find a quantitative estimate of the Human Development Stages by defining and calculating the specific age groups and subgroups. Early and late adolescence stages, first, second and third stages of adult development are estimated along with the early, middle and old age groups and subgroups. We briefly present the first exit time theory used to find the health state function of a population and then we give the details of the new theoretical approach with the appropriate applications to support and validate the theoretical assumptions. Our approach is useful for people working in several scientific fields and especially in medicine, biology, anthropology, psychology, gerontology, probability and statistics. The results are connected with the speed and acceleration of the deterioration of the human organism during age as a consequence of the changes in the first, second and third differences of the Health State Function and of the Deterioration Function.

**Introduction: The Health State Function Theory**

The advances in stochastic theory and the resulting the first exit time or hitting time theories gave rise to the development a new theoretical and applied approach for estimating the health state function (HSF) of a population and finding various very important parameters useful in many applications including the life expectancy and the healthy life expectancy. In this study we propose quantitative methods of estimating the human development age groups based on the findings of the health state function theory. We briefly present the first exit time theory used to find the health state function of a population (HSF) and then we give the details of the new theoretical approach with the appropriate applications to support and validate the theoretical assumptions.

In developing the health state function theory the original problem was to give a definition of the health state. It is not an easy task to find an exact and precise definition of health of an individual. Even more it is relatively difficult to define and measure the health state of a person. However, as for very many language terms introduced centuries or thousand of years ago the approval and use from the public is straightforward. According to the World Health Organization (WHO) definition in 1946 health is: "a state of complete physical, mental, and social well-being and not merely the absence of disease or infirmity."

When we started working on the paper of 1995 (Janssen, J. and Skiadas, C. H., Dynamic Modelling of Life-Table Data, *Applied Stochastic Models and Data Analysis*, vol. 11, No 1, 35-49, 1995) it was clear that a definition of health was very important for our study. As we where ready to use the relatively new tools of stochastic theory, that it was essential it was to define the "health state" of an individual in the course of time as a term related the health. We called $S_t$ the health state at a time $t$. Accordingly, when the time is replaced by the age $x$, the health state is denoted by $S_x$. We soon get to the conclusion that an estimation of the health state was very difficult following the luck of a precise definition and measure of health of the individual. Health and health state are "stochastic variables" that is quantities which present slow, sharp or even dramatic changes over time (see figure 1 Graph A for few stochastic paths of individuals; the light red lines). However, for the modern stochastic theory there is no need to find the course of the health state for every individual. The task is to find the summation of many health states of the individuals that is to define and find the "Health State of a Population". In this case the health state of a population could be a smooth function over time or better over age group; a function which can be approximated by a relatively simple mathematical form expressed by the dark red curve in Graph A of figure 1.

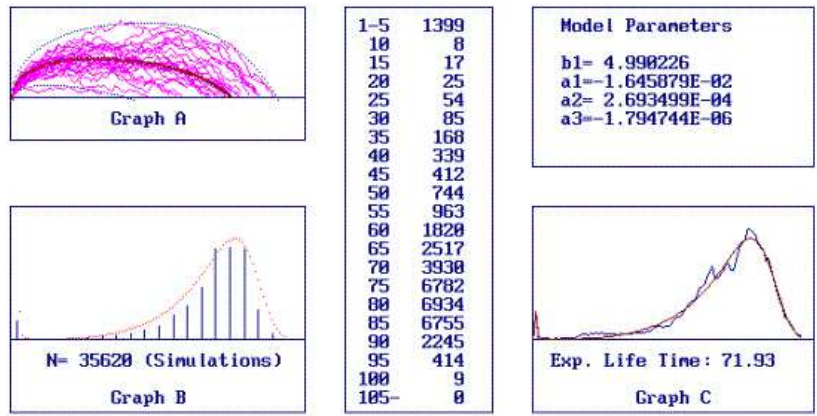

Fig. 1. Health State Function, model fitting, stochastic paths and
stochastic simulation for Greece, males (1992).

The next problem we had was the selection of data. The selection of data
related to the health or the health state was a difficult or impossible task
due to the luck of systematically collected data sets for long periods of
time along with the previously mentioned problems with the definition
of health and the health state. Especially when trying to collect survey
data from various countries and population groups the main problem is
on how individuals respond on questions related to their health and
health state. Cultural, economic or societal issues take particular
importance in answering the related questions.

The only reliable data sets are the population and death data selected for
centuries by the bureau of the census of every country. These data sets
almost precisely reply to the question Alive or Dead. "Alive" represents
the number of the members of a population for an age group and "Dead"
represents the number of dead of the members of the same group of
population at the same time period. We assume that death is the cause of
reaching a very law level for the health state while alive is a higher level
health status. The resulting death time series include vital information for
the health state of a population as definitely include the picture of the
population members reaching a law health level at specific years of age
as a distribution over years of age. This is called the hitting time or first
exit time distribution in terms of the stochastic theory. There are many
stochastic theory technicalities related on how to find the health state
function (*the cause*) from *the result* that is the distribution function (the
death distribution) but finally we arrived to a final solution in the paper
of 1995. The application was done for data in France and Belgium thus
giving the form for the health state of the population in these countries.

An application for Greece (males 1992) is illustrated in Figure 1. The first exit time model proposed was applied to the data by using a nonlinear regression analysis algorithm. Graph C of figure 1 presents the data (blue line) and fit curve (red line) and the estimated expected life time. In top right the estimated parameter values are given. The resulting Health State Function is estimated by using these parameters and is the basis of Graph A of figure 1(dark red line). Several health state random realizations for individuals are presented (light red lines). End of life results when these paths approach the bar of the minimum health.

The next step is to verify the validity of our theory by means of solving the inverse problem that is: to find the death distribution function when the health state function is known or estimated, by performing stochastic simulations. 35620 stochastic path realizations are done and the results are summarized in a Table in the middle of figure 1 in 5 year age groups. The related graph is Graph B of figure 1 where the simulation bars approach the death probability distribution (dotted line) relatively well.

By using the results of the theory which we have applied we arrive to the very important point that is to find the form of the health state function of a population by only death results. Of course we normalize the death distribution by using the population distribution. In other words we take into account the number of the population per age group. In the case of a "stable population" in terms of Keyfitz (2005) we do not need such adaptation and only the number of deaths per age is needed. We have also proposed a method of finding the death distribution $g(x)$ by using the mortality data usually referred as $\mu(x)$. For more information and theoretical and applied details see our references (Skiadas 2011a,b, 2012 and Skiadas & Skiadas 2010, 2011 and 2012a,b)

A next approach came only recently (Skiadas 2012) by solving the inverse problem on how to find the health state function of a population over age from the death distribution expressed by the death probability density function presented in figure 2A. This is found after dividing the death data of a population per year of age by the sum of deaths for all age years thus providing a data set with sum equal to unity. In this way it is easy to compare death data from various time periods or from different population groups. The case in our example is for males in USA the year 2000 and both the original data points and the fit curve by using our model are present. Figure 2B illustrates the health state function resulting from the same data and from the fit curve and Figure 2C presents the heath state change (first derivative) of the USA males in 2000 from data and curve fit. The continuous line in figure 2C refers to the first difference (the derivative) of the health state function.

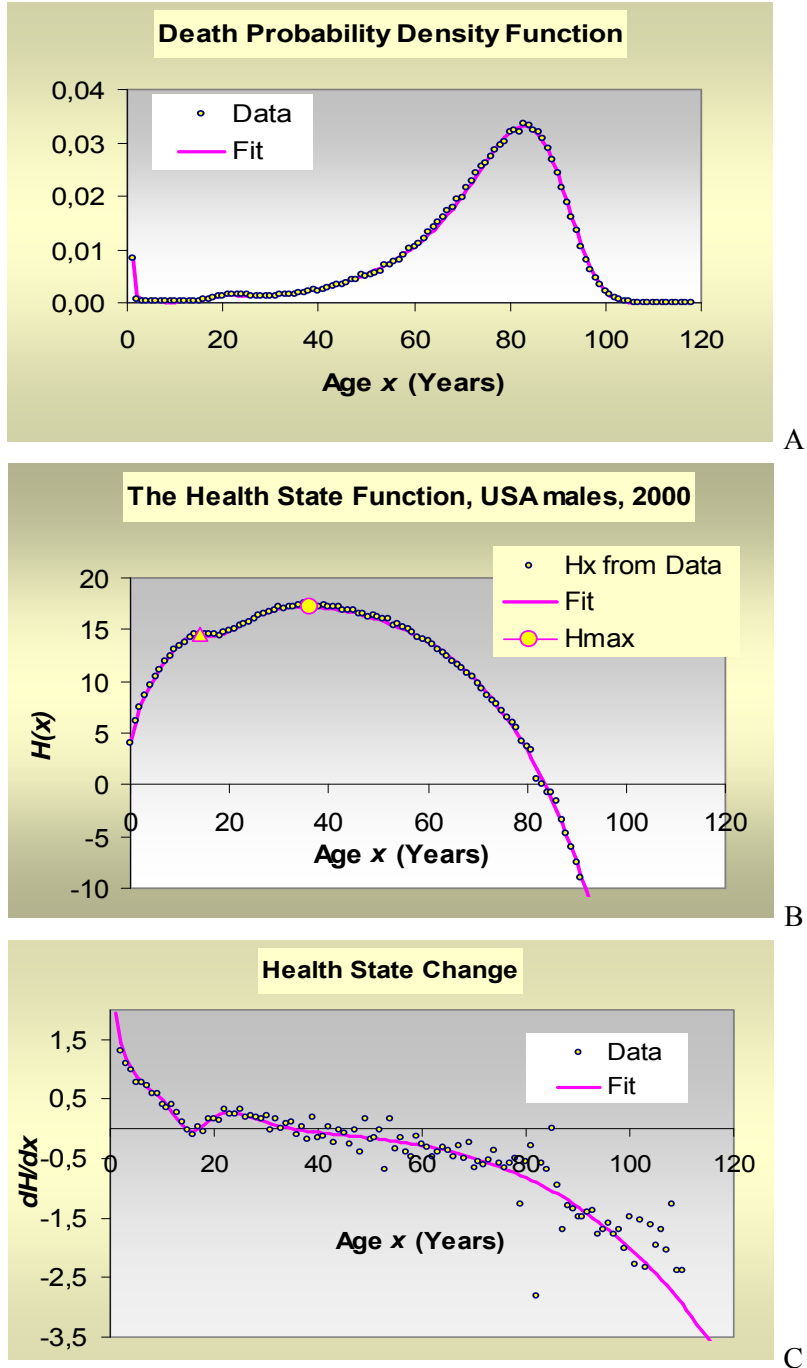

Fig. 2. The Health State Change of a Population

**Human Development Stages based on the Health State Function**

The Health State Theory and the resulting models and applications provide the opportunity to explore the development of the early and middle stages of the human life span. This is feasible by estimating the first and second variation of the health state changes or in mathematical notation to estimate the first and second derivative of the health state function (see the second variation in Figure 3).

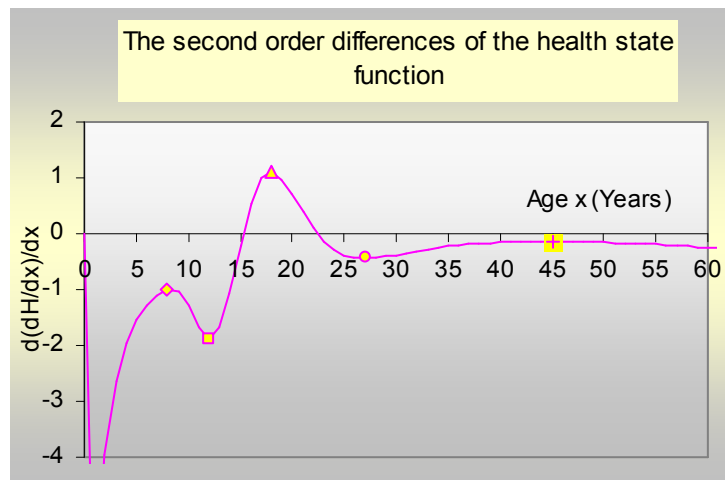

Fig. 3. The second order differences of the health state function.

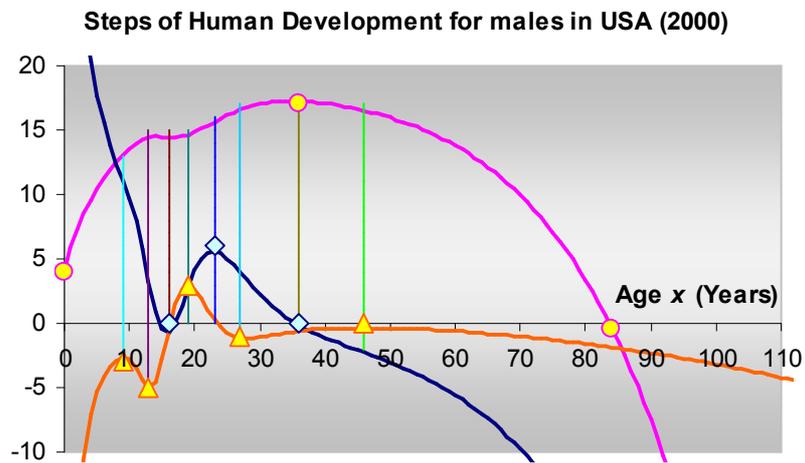

Fig. 4. Health state function, first and second order differences.

There are three main graphs to analyze the human development by based on the health state function of a population: the health state graph, the first difference graph and the second difference graph. Each graph provides characteristic milestones of the human development (see figures 4 and 5 for males and females in USA the year 2000).

For males in USA in 2000 the health state graph (Magenta curve in figure 4) has a starting point at zero age (health level: 4.00) a maximum health age at 36 years of age (17.22) and a zero health state at 84 years of age. After this year of age the health is in a critical and supercritical condition. The death probability is growing exponentially.

The graph of the first differences expresses the speed of health state change (Blue curve in figure 4). The first part of this graph starts from very high positive values at birth and fast decreases at a value close to zero (-0.75) at 16 years of age. It follows a growing part to a maximum (5.6) at 23 years of age. Then the decline is continuous passing from positive to negative values at 36 years of age corresponding to the health state maximum.

The acceleration of the health state is illustrated in the graph of the second differences (Orange curve in figure 4) and provides characteristic values for the years of age 9, 13, 19, 27 and 46. All these time steps correspond to changes of the slop of the speed of health change (the first differences graph) thus indicating characteristic changes in human development. The last one, that is 46 years of age, corresponds to the starting point of faster and faster growing negative growth of the speed of health changes. This means that is the starting point of the deterioration of the human organism via the aging process. Then the decline of the health state is a merely continuous process in terms of the health state function. Instead the first part of the life span from zero age until the maximum health state (36 years for males in USA the year 2000) is extremely interesting indicating on how the human complex organism is developed.

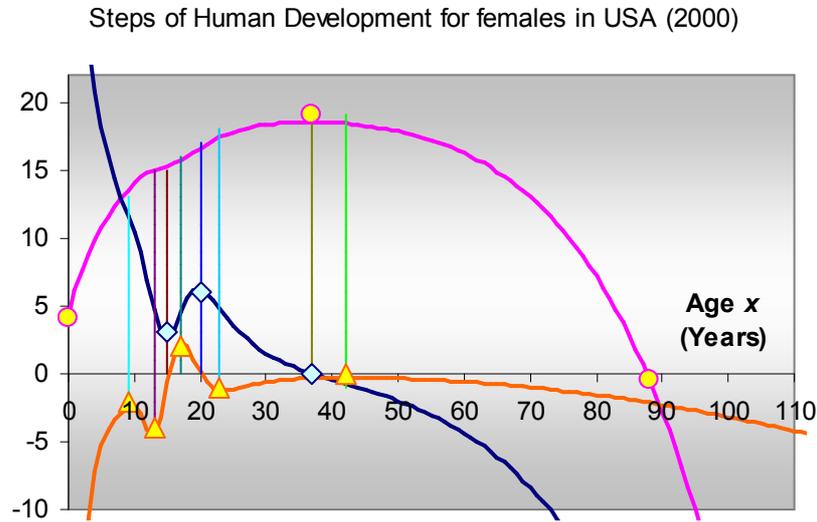

Fig. 5. Health state function, first and second order differences.

The results of the application for USA males, the year 2000 are summarized in Tables I and II and illustrated in the previous Figures 4 and 5. In the same Tables related results are given for UK, Australia, Canada, Germany, France, Italy and Japan for males and females in 2000. As it was expected there are very many similarities especially for the countries USA, UK, Australia and Canada. Our estimates provide useful information for the Pre-Adolescence, Early-Adolescence, Late-Adolescence age years ranging from 9 (minimum) to 21 (maximum) years of age for males and 8 (minimum) to 19 (maximum) years of age for females and followed by the First, Second and Third Stage of Adult Development starting from 19 (minimum) to 37 (maximum) years of age for males and from 16 (minimum) to 41 (maximum) years of age for females. The next age groups refer to Early Middle ages from 33 (minimum) to 48 (maximum) years of age for males and 37 (minimum) to 46 (maximum) years of age for females, Middle+Old ages from 43 (minimum) to 85 (maximum) years of age for males and 41 (minimum) to 91 (maximum) years of age for females and Very Old ages from 83+ (minimum) years of age for males and 87+ (minimum) years of age for females.

TABLE I

| Males 2000 | Human Development Age Groups by based on the Health State Function | | | | | | | | |
|---|---|---|---|---|---|---|---|---|---|
| Country | Pre Adoles cence | Early Adoles cence | Late Adolesce nce | First Stage of Adult Develop ment | Second Stage of Adult Develop ment | Third Stage of Adult Develop ment | Early Middle Ages | Middle+ Old Ages | Very Old Ages |
| USA | 9-13 | 13-16 | 16-19 | 19-23 | 23-27 | 27-36 | 36-46 | 46-84 | 84- |
| UK | 9-12 | 12-17 | 17-20 | 20-25 | 25-32 | 32-37 | 37-48 | 48-83 | 83- |
| Australi a | 9-12 | 12-17 | 17-21 | 21-27 | 27-34 | 34-41 | 41-48 | 48-85 | 85- |
| Canada | 8-12 | 12-16 | 16-19 | 19-24 | 24-29 | 29-37 | 37-47 | 47-85 | 85- |
| Germa ny | 12-15 | 15-17 | 17-19 | 19-23 | 23-26 | 26-34 | 34-43 | 43-83 | 83- |
| France | 11-15 | 15-17 | 17-19 | 19-22 | 22-26 | 26-33 | 33-43 | 43-85 | 85- |
| Italy | 9-12 | 12-17 | 17-20 | 20-24 | 24-30 | 30-37 | 37-47 | 47-84 | 84- |
| Japan | 11-14 | 14-16 | 16-19 | 19-23 | 23-26 | 26-36 | 36-43 | 43-86 | 86- |

TABLE II

| Females 2000 | Human Development Age Groups by based on the Health State Function | | | | | | | | |
|---|---|---|---|---|---|---|---|---|---|
| Country | Pre Adoles cence | Early Adoles cence | Late Adolesce nce | First Stage of Adult Develop ment | Second Stage of Adult Develop ment | Third Stage of Adult Develop ment | Early Middle Ages | Middle +Old Ages | Very Old Ages |
| USA | 9-13 | 13-15 | 15-17 | 17-20 | 20-23 | 23-37 | 37-42 | 42-88 | 88- |
| UK | 10-13 | 13-16 | 16-18 | 18-21 | 21-25 | 25-38 | 38-43 | 43-87 | 87- |
| Australia | 10-13 | 13-16 | 16-18 | 18-21 | 21-25 | 25-41 | 41-43 | 43-89 | 89- |
| Canada | 8-10 | 10-14 | 14-16 | 16-20 | 20-24 | 24-40 | 40-43 | 43-89 | 89- |
| Germany | 10-13 | 13-16 | 16-18 | 18-21 | 21-24 | 24-40 | 40-41 | 41-88 | 88- |
| France | 9-11 | 11-15 | 15-16 | 16-19 | 19-23 | 23-41 | 41-45 | 45-90 | 90- |
| Italy | 8-12 | 12-16 | 16-19 | 19-24 | 24-30 | 30-41 | 41-42 | 42-89 | 89- |
| Japan | 12-15 | 15-17 | 17-19 | 19-21 | 21-24 | 24-40 | 40-46 | 46-91 | 91- |

## Human Development Stages based on the Death Probabilities

Another point is to explore the steps of human development by using the traditional method of computing the death probabilities as a function of age $\mu(x)$ of a population. The estimation of the death probabilities is the basis of the classical method of constructing life tables in actuarial science. The death probability (the probability of dying within a specific age year) is estimated by dividing the number of deaths in a specific year of age by the living population in the same year of age. In general, the probability of death increases over time thus growing in the opposite direction of the health state of the population. However, a mortality method of defining the age groups by based on the death probabilities may give interesting findings compared with those provided by the health state method. The resulting graphs of an application in USA males in 2000 are presented in figure 6. The graphs represent the death probabilities $\mu(x)$, the fist differences $\mu(x)$' and the second differences $\mu(x)$''. The characteristic points estimated are at 9, 14, 17, 19, 21, 23, 27 and 34 years of age. The values are very close or identical to the estimates based on the health state function.

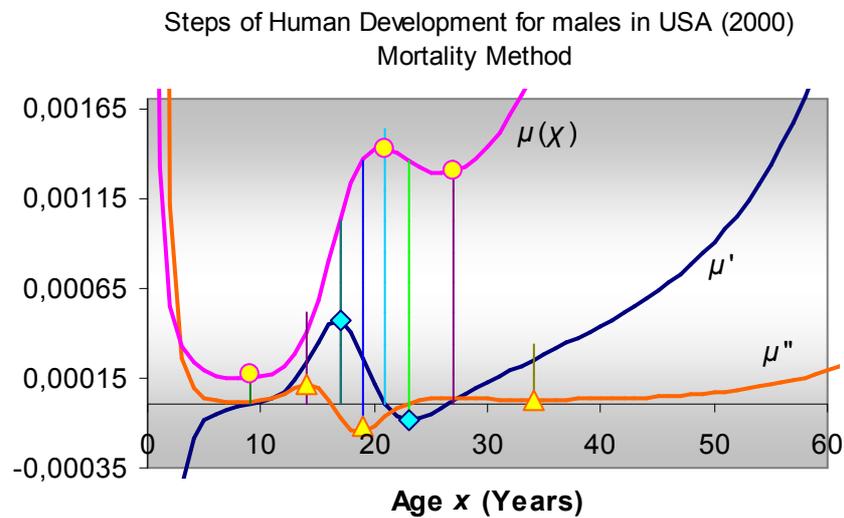

Fig. 6. Mortality function and first and second order differences.

**An Application to Disease Data in USA**

There are two shortcomings by using the death probabilities method. While the year with the maximum health is estimated with a small decline of two years from the health state method estimates (34 years instead of 36), the other estimate related to the early middle ages period at 46 years of age is totally missing. This is a very important age step connected with the acceleration of the decline of the health state. That is why it is the starting point of diseases like prostate cancer for males and uterine cancer with females.

Cancer and other data from USA and UK support this argument. Table III includes related data for USA in 2007.

TABLE III

| Cause of death | Under 1 year | 1 to 4 years | 5 to 14 years | 15 to 24 years | 25 to 34 years | 35 to 44 years | 45 to 54 years | 55 to 64 years | 65 to 74 years | 75 to 84 years | 85 years and over | Sum |
|---|---|---|---|---|---|---|---|---|---|---|---|---|
| | | | | | USA 2007 | | | | | | | |
| Malignant neoplasm of larynx | 0 | 0 | 0 | 2 | 2 | 55 | 450 | 932 | 989 | 871 | 333 | 3,634 |
| Malignant neoplasms of corpus uteri and uterus, part unspecified | 0 | 0 | 0 | 2 | 31 | 168 | 584 | 1,583 | 2,021 | 1,960 | 1,107 | 7,456 |
| Malignant neoplasm of prostate | 0 | 1 | 0 | 1 | 1 | 21 | 428 | 2,271 | 5,716 | 11,257 | 9,397 | 29,093 |
| Malignant neoplasm of bladder | 1 | 0 | 0 | 0 | 7 | 93 | 570 | 1,564 | 2,817 | 5,009 | 3,782 | 13,843 |
| Multiple myeloma and immunoproliferative neoplasms | 0 | 0 | 0 | 2 | 7 | 155 | 705 | 1,788 | 2,917 | 3,863 | 1,879 | 11,306 |
| Parkinson's disease | 0 | 0 | 0 | 2 | 2 | 12 | 60 | 396 | 2,310 | 9,363 | 7,911 | 20,056 |
| Alzheimer's disease | 0 | 0 | 0 | 0 | 1 | 8 | 95 | 728 | 3,984 | 23,009 | 46,804 | 74,629 |
| Other acute ischemic heart diseases | 2 | 0 | 0 | 3 | 17 | 109 | 376 | 679 | 740 | 1,021 | 1,145 | 4,092 |
| Atherosclerosis | 1 | 0 | 0 | 2 | 1 | 27 | 134 | 350 | 829 | 2,238 | 4,590 | 8,232 |
| Emphysema | 3 | 0 | 1 | 1 | 10 | 60 | 486 | 1,590 | 3,294 | 4,835 | 2,509 | 12,789 |
| Infections of kidney | 5 | 1 | 1 | 8 | 6 | 30 | 59 | 64 | 91 | 170 | 193 | 628 |
| Hyperplasia of prostate | 0 | 0 | 0 | 0 | 0 | 1 | 1 | 12 | 47 | 147 | 283 | 491 |
| Inflammatory diseases of female pelvic organs | 4 | 0 | 0 | 1 | 5 | 6 | 14 | 19 | 22 | 21 | 24 | 116 |
| Total | 16 | 2 | 2 | 24 | 90 | 745 | 3,982 | 11,976 | 25,777 | 63,814 | 79,957 | 186,365 |

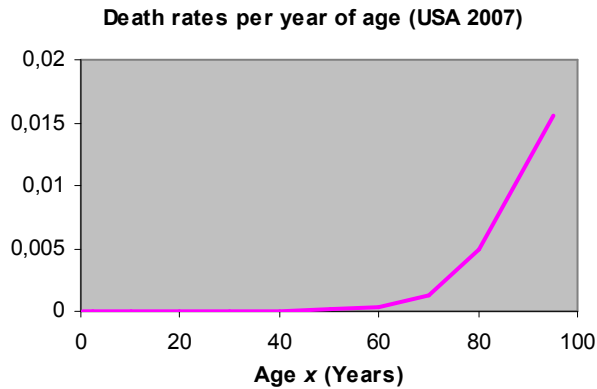

Fig. 7. Death rates for USA 2007 based on the TABLE III data (total)

The death rates per year of age for USA the year 2007 based on the previous Table III are presented in the figure 7. The cases selected from the USA data base include diseases which start from the age of the maximum health state. Significant values of the order of 0.4% of the total causes appear for the age group 35 to 44 years. Instead the previous years cases (from 0 to 34 years) account only for the 0.072% of the total cases. After the age group 35 to 44 the death rates follow an exponential trend. The estimates based on the USA 2007 data for males and females provide the following age groups (see TABLE IV). The decline starts after the maximum health state level at 36 years of age but the main causes start from 45 years of age a key point characterizing the onset of middle and old ages.

TABLE IV

| | Human Development Age Groups by based on the Health State Function | | | | | | | | |
|---|---|---|---|---|---|---|---|---|---|
| | Pre Adolescence | Early Adolescence | Late Adolescence | First Stage of Adult Development | Second Stage of Adult Development | Third Stage of Adult Development | Early Middle Ages | Middle+ Old Ages | Very Old Ages |
| Total USA 2007 | 10-14 | 14-17 | 17-19 | 19-23 | 23-28 | 28-36 | 36-45 | 45-88 | 88- |

## Human Development Age Groups Based on the Deterioration Function

The age groups after 40-45 years of age until 85-90 years where the zero health state level appears is a quite large period of the life time and it is worth noting to make separations in specific age groups or subgroups. For this case we can use the deterioration function (Det) and the first (Det'), second (Det'') and third (Det''') differences (derivatives) of this function in order to define and characterize these groups. The deterioration function is a measure of the curvature of the health state function $H(x)$ and thus it can be used to find how and how fast the HSF is changing. The formula measuring the curvature requires the calculation of the first and second derivatives of the health state function and has the form Det=|H''|/(1+H'$^2$)$^{3/2}$. The deterioration function is a very good measure of the decline of the human health state in the time course. Furthermore we can estimate the total influence of the deterioration process (TDET) by the following formula:

$$TDET = \int_{x(\min Det)}^{x} Det(s)ds \approx \sum_{x(\min Det)}^{x} Det(s)$$

Where the starting year denoted by $x_{minDet}$ is at the minimum deterioration age and the final year denoted by $x = x_{110}$ is at the 110 year of age.

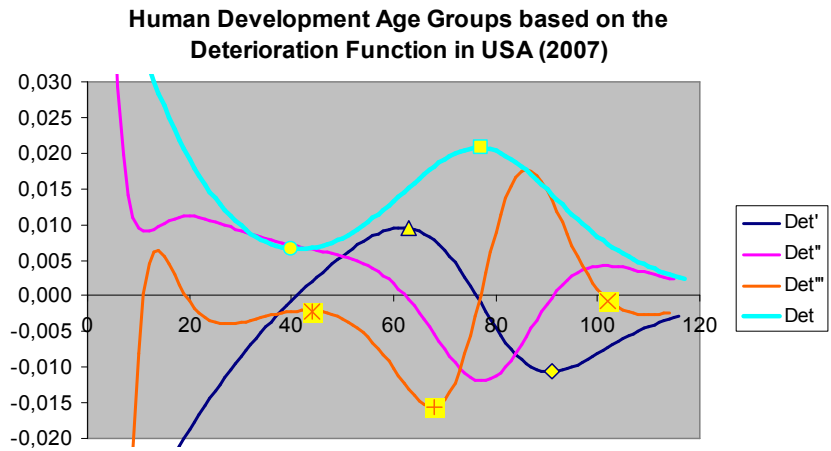

Fig. 8. The Deterioration function and the three differences of this function

Figure 8 illustrates the deterioration function and the first, second and third differences (derivatives) for USA in 2007 (males). The deterioration function (light bleu line) starts from very high positive values at birth and declines until age 41 where the minimum deterioration age appears, it grows to a maximum at 76 years and then continuously declines. It is worth noting that the first part of this function from birth until the minimum corresponds to the adult development stage. The part from the minimum until the late age years corresponds to the human deterioration life period. The first difference (blue line) has maximum and minimum corresponding to 61 and 90 years of age respectively. These age year steps represent the points where the exponential growth or decline of the deterioration function slows down. The first difference expresses the speed of the deterioration. The speed is positive for the first period of the deterioration (from 41 to 76 years of age) and then is negative for all the period of over 76 years of age. The second difference (magenta line) of the deterioration function provides an estimate of changes at 101 years of age. The second difference expresses the acceleration of the deterioration function. The acceleration is positive until 61 years of age then is negative until 90 years of age and then is positive for the rest of the life span. The third difference (orange line) of the deterioration function provides three characteristic points at 67, 85 and 109 years of age.

The estimated values for USA males and females for 2000 are summarized in Table V. In the same Table the Total Deterioration (TDET) is estimated in the last year of each age group along with the percentage of the total deterioration in the last year of each age group. The estimates for both males and females are very close each other especially as regards the % total deterioration. According to these findings the total deterioration is 24.0% and 23.8% at 60 years of age for males and females respectively. The related values are 33.2 and 32.5 at 65 and 66 years of age for males and females. Total deterioration close to 50% is estimated at the maximum deterioration age at 74 (TDET=52.6%) and 75 (TDET=53.1%) years of age for males and females respectively. The total deterioration is close to ¾ of the final, that is 74.4 for males and 74.8 for females in age years 84 (males) and 85 (females). In 100 years of age the total deterioration is 95.0 for males and 94.3 for females whereas this is 99.2% for males and 99.1% for females at 108 years of age while we have selected TDET=100% at 110 years of age. This selection was done because the data provided by the Human Mortality Database cover the range from 0 to 110 years of age.

TABLE V

| Human Development Age Groups by based on the Deterioration Function, Total Deterioration and % Deterioration in USA (2000) | | | | | | | | |
|---|---|---|---|---|---|---|---|---|
| Age Groups Total Deterior* %Total Det** | First Deterioration Stage | Second Deterioration Stage | Light disabilities stage | Moderate & Severe disabilities stage | Old Ages (many disabilities) | Very old ages (critical health) | Critical ages (very critical health) | Highly Critical ages |
| Age Groups Total Det Males %Tot Det | 41-60 0.21 24.0 | 60-65 0.29 33.2 | 65-74 0.46 52.6 | 74-84 0.65 74.4 | 84-89 0.72 82.4 | 89-100 0.83 95.0 | 100-108 0.867 99.2 | 108- 0.874 100 |
| Age Groups Total Det Females %TDet | 40-61 0.22 23.8 | 61-66 0.30 32.5 | 66-75 0.49 53.1 | 75-85 0.69 74.8 | 85-89 0.75 81.3 | 89-100 0.87 94.3 | 100-108 0.915 99.1 | 108- 0.923 100 |
| * The Total Deterioration is estimated for the final year of the age group** The % Total Deterioration is based on a 100% total deterioration in the year 110 | | | | | | | | |

An application to Sweden (males) gave interesting results regarding the development of the Deterioration and the Total Deterioration in the time course. Sweden provides reliable death and population data for the last two and a half centuries. In our study we explore data from 1800, 1900, 2000 and 2010 for males and the results are presented in Table VI. The main finding is that the Total Deterioration Percentage in every group is relatively stable for all this two hundred years period. Instead the years of age for the various groups are moved to higher ages supporting the argument of an improvement of the health state during a large period of time mainly due to the advances in medicine and social and economic welfare of the country. Figure 9 illustrates these findings for Sweden. The gap between age groups is smaller as we move to higher age groups. Following the results presented in Table V and VI we can form the following age groups already presented in these Tables. The First Deterioration age group is at 20%-25% TDET, the Second Deterioration age group is at 30%-33% TDET. It follows a Light Disability Stage group at 50%-55% TDET, a Moderate and Severe Disabilities age group at 70%-75% TDET and an Old Ages group with many disabilities at 80%+- TDET. Then the next two periods refer to Critical Health state with TDET at 90%-95% and Very Critical Health with TDET at 97%-99% years of age. The final group refers to Highly Critical ages with TDET close to 100%.

Accordingly the Table VII for males in year 2000 for USA, Australia, UK, Canada, Germany, France, Italy and Japan is constructed. The findings indicate a similar behavior for the age groups proposed supporting our theory and method for constructing these age groups based on death and population data.

TABLE VI

| Human Development Age Groups by based on the Deterioration Function, Total Deterioration and % Deterioration for Males in Sweden (1800-2010) | | | | | | | | |
|---|---|---|---|---|---|---|---|---|
| Age Groups Total Deterior* %Total Det** | First Deterioration Stage | Second Deterioration Stage | Light disabilities stage | Moderate & Severe disabilities stage | Old Ages (many disabilities) | Very old ages (critical health) | Critical ages (very critical health) | Highly Critical ages |
| Age Groups Total Det 1800 % Tot Det | 31-49 0.23 24.0 | 49-54 0.31 32.3 | 54-64 0.50 52.1 | 64-75 0.69 71.9 | 75-80 0.77 80.2 | 80-93 0.89 92.7 | 93-102 0.93 96.9 | 102- 0.96 100 |
| Age Groups Total Det 1900 % Tot Det | 34-55 0.22 21.8 | 55-60 0.32 31.7 | 60-68 0.52 51.5 | 68-76 0.72 71.3 | 76-80 0.80 79.2 | 80-89 0.91 90.1 | 89-96 0.96 95.0 | 96- 1.01 100 |
| Age Groups Total Det 2000 % Tot Det | 40-58 0.21 23.2 | 58-63 0.30 33.1 | 63-72 0.48 52.9 | 72-81 0.66 72.8 | 81-86 0.74 81.6 | 86-97 0.85 93.7 | 97-105 0.89 98.1 | 105- 0.907 100 |
| Age Groups Total Det 2010 % Tot Det | 40-59 0.19 20.3 | 59-65 0.30 32.0 | 65-73 0.49 52.3 | 73-81 0.67 71.5 | 81-87 0.77 82.2 | 87-95 0.87 92.8 | 95-102 0.91 97.1 | 102- 0.937 100 |
| * The Total Deterioration is estimated for the final year of the age group ** The % Total Deterioration is based on a 100% total deterioration in the year 110 | | | | | | | | |

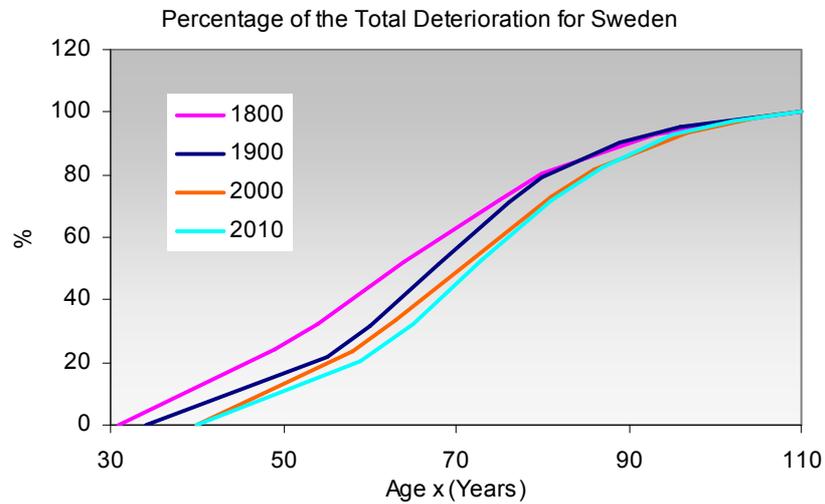

Fig. 9. The effect of the Total Deterioration in Sweden (males)

TABLE VII

| Human Development Age Groups by based on the Deterioration Function for Males (2000) in several Countries | | | | | | | |
|---|---|---|---|---|---|---|---|
| Age Groups | First Deterioration Stage | Second Deterioration Stage | Light disabilities stage | Moderate & Severe disabilities stage | Old Ages (many disabilities) | Very old ages (critical health) | Critical ages (very critical health) |
| USA | 41-60 | 60-65 | 65-74 | 74-84 | 84-89 | 89-100 | 100- |
| UK | 41-56 | 56-60 | 60-70 | 70-83 | 83-88 | 88-101 | 101- |
| Australia | 39-57 | 57-62 | 62-73 | 73-82 | 82-87 | 87-99 | 99- |
| Canada | 40-57 | 57-63 | 63-72 | 72-84 | 84-88 | 88-100 | 100- |
| Germany | 40-57 | 57-62 | 62-72 | 72-83 | 83-88 | 88-100 | 100- |
| France | 40-60 | 60-65 | 65-74 | 74-84 | 84-88 | 88-98 | 98- |
| Italy | 40-57 | 57-62 | 62-71 | 71-83 | 83-88 | 88-101 | 101- |
| Japan | 40-58 | 58-64 | 64-73 | 73-84 | 84-88 | 88-100 | 100- |

**Further Analysis and Quantification**

It should be noted that many studies in last decades emphasize on the estimation of the healthy life expectancy and the loss of healthy life years by light, moderate or severe disability causes. These studies provide tables presenting the healthy life expectancy and the loss of healthy life years in various countries. They give important information in countries and government officials, social and economic agencies, insurance companies and actuaries in order to adapt their plans according to the new findings and to forecast the future.

However, the healthy life expectancy as the life expectancy are mainly statistical estimates and provide a part of the hidden information in life table data. Instead with the method proposed we extract vital information for the health condition of various age groups by using death and population data. The simplest is to obtain the related data from an international data base as of the World Health Organization or the Human Mortality Database (HMD) used in this study. We use the Deaths 1x1 and the Population 1x1 data for USA, Sweden and the other countries from HMD and we use the program [Excel File of the SKI-6-Parameters_22_06_2012_Program_(zip_format_7Mb)](...) which we have developed in Excel, from the website [http://www.cmsim.net/id24.html](http://www.cmsim.net/id24.html) . By inserting the death and population data for a country for a specific year in columns P and Q in the Excel program, the Health State Function and the Deterioration Function are calculated. Another method is to insert in column R of the Excel program the values for $\mu(x)$ included in files as the life table data mltper 1x1 for males in the HMD or to obtain the death distribution from a Life Table, insert the data directly in column Q and set the indicator equal to 0 in cell AB24 of the Excel program. When we use the life table mltper 1x1 for males from the HMD Database the death distribution is given in column $d_x$ for the age years from 0 to 110.

Our results should also be useful to scientists working in the fields of human development and the related theories as are the Erikson's theory of personality and the proposed "Erikson's stages of psychosocial development", the Piaget, Sullivan and others in providing a quantitative measure of the proposed age groups of these theories. A first approach is given in TABLE VIII from results estimated from males and females data in USA (2000). The pre-adolescence, early-adolescence and late-adolescence periods are defined thus quantifying the age groups suggested by the related theories by using the Health State Function

Theory. The results from the deterioration function approach provide a better classification for adulthood period (see Tables V, VI and VII).

TABLE VIII

| Methods | Human Development Age Groups by based on the Health State Function | | | | | | | | |
|---|---|---|---|---|---|---|---|---|---|
| | Pre Adolescence | Early Adolescence | Late Adolescence | First Stage of Adult Development | Second Stage of Adult Development | Third Stage of Adult Development | Early Middle Ages | Middle+ Old Ages | Very Old Ages |
| Erikson's stages | Competence | Adolescence | | Early adulthood | | *Adulthood (27-64), Old Ages (64-) | | | |
| Sullivan's | Pre-adolescence | Early Adolescence | Late Adolescence | | | | | | |
| Piaget's | Concrete Operations | Formal operational stage | | | | | | | |
| Males USA 2000 | 9-13 | 13-16 | 16-19 | 19-23 | 23-27 | 27-36 | 36-46 | 46-84 | 84- |
| Females USA 2000 | 9-13 | 13-15 | 15-17 | 17-20 | 20-23 | 23-37 | 37-42 | 42-89 | 89- |
| * The correct group formation according to Erikson is found by using the deterioration function group classification. In this case there appear age groups close to 65 years (see Tables V, VI and VII for validation) | | | | | | | | | |

## Conclusions

The method of estimating the health state function of a population the deterioration function and the first and second and third differences (derivatives) provided a method to estimate the Human Development Age Groups. The resulting applications in various countries supported the argument of characteristic age groups. The proposed human development age group selection will be useful in scientific fields like medicine, sociology, biology, anthropology, psychology, gerontology, probability and statistics and many others. We provide a method to quantify the human development age groups thus giving reliable tools for further studies.

## References

J. Janssen and C. H. Skiadas, Dynamic modelling of life-table data, *Applied Stochastic Models and Data Analysis*, **11**, 1, 35-49 (1995).

N. Keyfitz and H. Caswell, *Applied Mathematical Demography*, 3rd ed., Springer (2005).

C. H. Skiadas, A Life Expectancy Study based on the Deterioration Function and an Application to Halley's Breslau Data, arXiv:1110.0130 [q-bio.PE] (1 Oct 2011), http://arxiv.org/ftp/arxiv/papers/1110/1110.0130.pdf .

C. H. Skiadas, Life Expectancy at Birth, Estimates and Forecasts in the Netherlands (Females), arXiv:1112.0796 [q-bio.PE] (4 Dec 2011), http://arxiv.org/ftp/arxiv/papers/1112/1112.0796.pdf

C. H. Skiadas, The Health State Function, the Force of Mortality and other Characteristics resulting from the First Exit Time Theory applied to Life Table Data, arXiv:1202.1581v1 [q-bio.PE] (8 Feb. 2012, http://arxiv.org/ftp/arxiv/papers/1202/1202.1581.pdf ).

C. Skiadas and C. H. Skiadas, Development, Simulation and Application of First Exit Time Densities to Life Table Data, *Communications in Statistics* 39, 444-451 (2010).

C. H. Skiadas and C. Skiadas, Properties of a Stochastic Model for Life Table Data: Exploring Life Expectancy Limits, *arXiv*:1101.1796v1 (10 Jan 2011), http://arxiv.org/ftp/arxiv/papers/1101/1101.1796.pdf .

C. H. Skiadas and C. Skiadas, Estimating the Healthy Life Expectancy from the Health State Function of a Population in Connection to the Life Expectancy at Birth, arXiv:1205.2919 [q-bio.PE] (14 May 2012), http://arxiv.org/ftp/arxiv/papers/1205/1205.2919.pdf .

C. H. Skiadas and C. Skiadas, A Method for Estimating the Total Loss of Healthy Life Years: Applications and Comparisons in UK and Scotland, arXiv:1212.4583 [q-bio.PE] (19 December 2012), http://arxiv.org/ftp/arxiv/papers/1212/1212.4583.pdf .